\documentclass[preprint,12pt]{elsarticle}
\usepackage{amsmath}
\usepackage{amsfonts}
\usepackage{amssymb}
\usepackage[utf8]{inputenc}
\usepackage[T2A]{fontenc}
\usepackage{bm}
\usepackage{graphicx}

\begin{document}

	\title{Davydov's soliton in an external alternating magnetic field} 
	
	\author{Larissa Brizhik}
	\address{Bogolyubov Institute for Theoretical Physics of the National Academy of Sciences of Ukraine \\
		Metrologichna Str., 14-b,  Kyiv, 03143, Ukraine	\\[0pt]
		\vspace*{0.25cm}
		brizhik@bitp.kiev.ua }
	
	\begin{abstract}
		{
The influence of an external oscillating in time magnetic field on the dynamics of the Davydov’s soliton is investigated. It is shown that it essentially depends not only on the amplitude and frequency of the magnetic field, but also on the field  orientation with respect to the molecular chain axis. The soliton velocity and phase are calculated. They are oscillating in time functions  with the frequency of the main harmonic, given by the external field frequency, and higher multiple harmonics. It is concluded that such complex effects of external time-depending magnetic fields on the dynamics of solitons modify the charge transport in low-dimensional molecular systems, which can affect functioning of the devices based on such systems. These results suggest also the physical mechanism of therapeutic effects of oscillating magnetic fields, based on the field influence on the dynamics of solitons which provide charge transport through biological macromolecules in the redox processes.
		}
		
	\end{abstract}
	
	\maketitle

\section{Introduction}

It is 50 years since the principal idea and the simplest model of the molecular solitons have been formulated by A. Davydov and N. Kislukha in their paper in October, 1973 issue of Physica Status Solidi (b) \cite{DavydovKislukha1}. We briefly remind here that such solitons  are formed in soft molecular chains due to the electron-phonon interaction, they  describe self-trapped states of quasi-particles (electrons, holes, molecular excitations, etc.) in the deformation potential of molecular chains \cite{Davydov}. This model has been developed initially  to resolve the so called crisis in bioenergetics, and in the following years it was further extended in numerous papers not only by Davydov and his school, but also by scientists from many other countries. The soliton model was generalized to describe  various biological processes and different low-dimensional systems, taking into account many other factors, such as more complex structure of molecular chains, energy dissipation, impact of temperature, presence of external fields, etc. (see \cite{Scott}). Alwyn Scott has introduced the term  "Davydov's soliton" for such nonlinear localized states of quasi-particles in low-dimensional molecular systems. Some of  the first review papers on this subject can be found in \cite{Scott,Brizhik-dyn} and references therein. Significant advances in the development of the theory of molecular solitons have been achieved during recent years, some of which were published in  the special issue “To the 110th birthday of Alexander Davydov” of the journal Low Temperature Physics (J. Low. Temp. Phys, \textbf{48} (2022)).  

Low-dimensional molecular systems include not only biological macromolecules, but also a large class of organic and inorganic systems, such as polydiacetylene, conducting polymers, some superconducting compounds and many other, in which electron-lattice interaction is sufficiently strong and leads to self-trapping of quasiparticles in soliton states, providing, thus, efficient mechanism of charge and energy transport on macroscopic distances. Some of these compounds are widely used in microelectronics, and  many expectations in the advances of novel technologies are based on using biological macromolecules or polymers in "Donor -- Polymer -- Acceptor" arrays (see  \cite{Brizhik-DAA} and references therein). In many circumstances these systems are exposed to external fields, and, in particular, to magnetic fields. Following the study of the impact of the external constant magnetic field on the Davydov's soliton  \cite{BrizhikMF1,BrizhikMF2}, in the present paper we study the case of oscillating in time magnetic field. The importance of this subject  is connected  not only with the wide use of  oscillating magnetic fields in modern technologies and home appliances, which can affect charge transport in technical devices, but also such fields can affect the redox processes in living organisms. Moreover, there are numerous therapies, based on the low-intensity magnetic fields, which turned out to be successful in non-invasive treatment of various diseases. Their successful use in modern medicine, obviously, requires deep understanding of the underlying physical mechanisms \cite{BrizhikArXiv,BrizhikFermiZavan,BrizhikFermi}. The peculiarity of our approach is that it is based on the nonlinear mechanism of charge transport along macromolecules in the redox processes which can act in addition to other linear mechanisms (see, e.g., \cite{McLeod}). 

The paper is organized as follows. In Section 2 we introduce the Hamiltonian of a three-dimensional system comprising molecular chains, in the presence of the magnetic field. In Section 3 we consider the case when magnetic field is parallel to molecular chains and in Section 4 when it is perpendicular. It will be shown 
that soliton dynamics is described by system of coupled nonlinear equations, which under certain approximations can be reduced to the modified nonlinear Schr\"odinger equation. We will solve these equations using the nonlinear perturbation theory,  based on the method of the inverse scattering problem. We will show that in the periodic magnetic field, parallel to chains, soliton dynamics consists of electron motion along the chain in a free soliton state (soliton in the absence of the field) and of the oscillatory movement in the transverse direction, described by harmonic oscillator functions. Much more rich dynamics of soliton occurs in the transverse magnetic field, when the speed, width and energy of solitons turn out to be oscillating functions of time  with frequency oscillations given by the external field frequency and its higher harmonics. 
Finally, in Conclusions we summarize the obtained results and discuss generalization of the approach, presented in Ref. \cite{BrizhikArXiv}, to explain the possible mechanism of the magnetic field impact  on the metabolism of living organisms via its influence on the soliton provided charge transport along macromolecules during the oxidation-reduction reactions, and, based on this, mechanism of the low-intensity alternating magnetic field therapies.

\section{Equations of motion}

To study soliton dynamics in the external magnetic field, we have to introduce briefly what is a soliton or electrosoliton, when we speak about an extra electron in a molecular chain; namely such a case we will consider below.  It is known that the simplest model of the Davydov's soliton describes the ground state of an electron  in a molecular chain with acoustical phonon branch at moderate values of the electron-lattice coupling constant. Such a system is described  by the Fr\"ohlich Hamiltonian
\begin{equation}
	H_F={{H}_{e}}+{{H}_{ph}}+{{H}_{el-ph}},
	\label{HF}
\end{equation}
where the terms ${H}_{e}$, ${H}_{ph}$ and ${H}_{e-ph}$ are the Hamiltonians of the electron, phonons and their interaction, respectively, \cite{Davydov}. It leads to the 
nonlinear system of discrete dynamical equations, which in the long-wave approximation can be transformed to the continuum equations
\begin{equation}
	\label{eqpsi1}
	i\hbar \frac{\partial }{\partial t}\psi (x, t)=\left[ \frac{p_x^2}{2m_x}-\sigma \rho(x,t)\right]\psi (x,t),
\end{equation}
\begin{equation}
		\label{eqrho1}
	\left( \frac{{{\partial }^{2}}}{\partial {{t}^{2}}}-{{V}_{0}}^{2}\frac{{{\partial }^{2}}}{\partial {{x}^{2}}} \right)\rho (x, t)+\frac{\sigma {{a}^{2}}}{M}\frac{{{\partial }^{2}}}{\partial {{x}^{2}}}{{\left| \psi (x, t) \right|}^{2}}=0,
\end{equation}
for the electron wavefunction ${\psi (x, t)}$  and lattice deformation ${\rho (x, t)}$. Here  $x$ is the coordinate along the chain,  $p_x=i\hbar \partial  /\partial x $ is the electron momentum,  ${\sigma}$ is the electron-lattice coupling constant,  ${a}$ is the lattice constant (distance between the units of a molecular chain), $m_x=\hbar ^2 /2Ja^2$ is the effective electron mass, which is determined by the exchange interaction $J$,  ${M}$ is the mass of a molecule,  ${{V}_{0}}$ is the sound velocity in the chain. The electron wavefunction determines the probability of the electron localization and, hence, is normalized to unity. 

 In the adiabatic approximation the solution of Eq. (\ref{eqrho1}) is 
\begin{equation}
	\label{rho}
	\rho(x,t)=\frac{ \sigma }{w(1-s^2)}|\psi (x,t)|^2,  
\end{equation}
and Eq. (\ref{eqpsi1}) is transformed to the nonlinear Schr\"odinger equation (NLS)
\begin{equation}
	\label{NLS}
	\left( i\hbar \frac{\partial }{\partial t}+Ja^2\frac{{{\partial }^{2}}}{\partial {{x}^{2}}}+2Jg |\psi (x,t)|^2 \right)\psi (x,t)=0,
\end{equation}
with the well-known soliton solution
\begin{equation}
	\label{psi-sol}
	\psi (x,t)=\psi _s(x,t)\equiv \frac{\sqrt{g} }{2\cosh\left[ g(x-x_0-Vt)/a \right]} \exp \left[ i({{k}_{x}}x-\omega t) \right].
\end{equation}
Here ${g}$ is the dimensionless nonlinearity constant
\begin{equation}
	\label{g}
	g=\frac{ \sigma ^2}{2Jw(1-s^2)}, \qquad s^2=\frac{V^2}{V_0^2},
\end{equation}
$V$ is soliton velocity, $w=MV_0^2/a^2$ is the chain elasticity, $k_x=m_xV/\hbar$ is the $x$-component  of the electron wave-vector, $\omega $ is the frequency determined by the eigen-energy of the soliton, $x_0$ is the soliton centre of mass position at the initial time moment $t=0$ which for simplicity can be set  equal to zero.

In the effective mass approximation, the total soliton energy and its effective mass $M_s$ are given by the expressions
\begin{equation}
	\label{E-sol}
	E_s(V)=E_s(0)+\frac{1}{2}M_sV^2, \qquad M_s \approx m_x\left(1+\frac{M\sigma ^4}{6w^3\hbar^2}\right)
\end{equation}

To study the influence of the magnetic field on solitons, we have to generalize our model to a three-dimensional case and to represent the electron wavefunction in the form
\begin{equation}
	\label{psi3D}
	\psi (\vec{r}, t)=\psi (x,t)\psi_{tr} (y,z,t).
\end{equation}

  The  electron momentum $ {\vec p} ({\vec r})=i\hbar\partial / \partial {\vec r} $ in the presence of the magnetic field has to be modified as follows
\begin{equation}
	\label{mom}
	{\vec p}({\vec r}) \rightarrow	 {\vec p}({\vec r}) - \frac{e}{c}{\vec A} 
\end{equation}
where  ${\vec A}$ is the magnetic field vector-potential  ${\vec B}=\rm{rot}\vec{{A}}$, $e$ is the  electron charge and $c$ is the speed of light. 

Therefore, the electron wavefunction in the external magnetic field is determined by the equation
\begin{equation}
	\label{psi}
	i\hbar \frac{\partial }{\partial t}\psi (\vec{r}, t)=H\psi (\vec{r}, t),
\end{equation}
with the Hamiltonian (see Eqs. (\ref{HF})-(\ref{mom}))  
\begin{equation}
	\label{H}
	H=\sum\limits_{\nu =1}^{3}{\left[ {{\left( {{p}_{\nu }}-\frac{e}{c}{{A}_{\nu }} \right)}^{2}}\frac{1}{2{{m}_{\nu }}}-\sigma \rho (x, t) \right]},
\end{equation}
where $\nu =1,2,3$ or, equivalently, $\nu=x,y,z$.

An arbitrary harmonic magnetic field ${\vec B}(t)=\vec {B}_0\cos (\omega t)$ can be expressed in the general case via its longitudinal and perpendicular to the chain orientation components. In the next two sections we will study these two cases  separately.
	
	\section{The case of the longitudinal magnetic field}
	
Let us first consider the case when the harmonic magnetic field is parallel to the molecular chain: 
\begin{equation}
	\label{B-par}
	\vec{B}^{(l)}(t)=({{B}_{0}}\cos \omega t, 0, 0),
\end{equation}	 
where $\omega $ is the field oscillation frequency (see also brief report \cite{TemchBrizhik}).	We can choose the following gauge invariance of the magnetic field 
\begin{equation}
	\label{A-par}
	\vec{A}^{(l)}=(0, -{{B}_{0}}z\cos \omega t, 0).
\end{equation}
Substituting it into Eq. (\ref{H}), we can re-write the Hamiltonian as the  sum of the two parts $H=H_s^{(l)}+H_{tr}^{(l)} $, where 
\begin{equation}
	\label{Hspar}
	H_{s}^{(l)}=-\frac{\hbar ^{2}}{2m_{x}}\frac{\partial ^{2}}{\partial x^2}-\sigma \rho (x, t)				
\end{equation}
and
\begin{equation}
	\label{Htr}       
	H_{tr}^{(l)} = \frac{1}{2 m_x}
	\biggl(
	-\hbar^2 \frac{\partial^2}{\partial y^2} - \frac{2i\hbar e}{c} B_0 z \cos \omega t \frac{\partial }{\partial y}
	+
	\frac{e^2}{c^2} B_0^2 z^2 \cos^2 \omega t
	\biggr)
	-\frac{\hbar^2}{2 m_z}\frac{\partial^2}{\partial z^2}.
\end{equation}
with ${{m}_{\nu}}$ being the components of the effective electron mass in the conduction band of the system. Therefore, the electron wavefunction (\ref{psi3D}) can be represented as the product
\begin{equation}
	\label{psil}
	\psi^{(l)} (\vec{r}, t)=\psi_s (x,t)\psi _{tr} (y,z,t).
\end{equation}
and, therefore, it is determined by the system of two equations, the first of which is the equation for $\psi_s(x,t)$, which coincides with the equation for a free soliton (\ref{eqpsi1}), 
and the second equation is
$$
	i\hbar \frac{\partial {{\psi }_{tr }}(y, z, t)}{\partial t}+\frac{{{\hbar }^{2}}}{2{{m}_{y}}}\frac{{{\partial }^{2}}{{\psi }_{tr }}(y, z, t)}{\partial {{y}^{2}}}
	+\frac{i\hbar e}{{{m}_{y}}c}{{B}_{0}}z\cos \omega t\frac{\partial {{\psi }_{tr}}(y, z, t)}{\partial y}-
$$
\begin{equation} 
	\label{psi-tr-par}	
	-\frac{{{e}^{2}}}{2{{m}_{y}}{{c}^{2}}}{{B}_{0}}^{2}{{z}^{2}}{{\cos }^{2}}\omega t{{\psi }_{tr }}(y, z, t)
	+\frac{{{\hbar }^{2}}}{2{{m}_{z}}}\frac{{{\partial }^{2}}{{\psi }_{tr }}(y, z, t)}{\partial {{z}^{2}}}=0.
\end{equation} 
The solution of this equation can be searched in the form of the plane wave in one direction (e.g., $y$) and an some function in the second direction, to be determined below:
\begin{equation}
	\label{psitr1}
	{{\psi }_{tr }}(y, z, t)=\frac{1}{\sqrt{{{l}_{y}}}}\varphi (z, t)\exp \left( i{{k}_{y}}y-i\frac{{{E}_{tr }}t}{\hbar } \right),
\end{equation}
where ${l}_{y}$ is a characteristic size of the system in the corresponding direction, which determines the normalization condition, and 
$$ {E}_{tr }= \frac{\hbar ^2{k}_{y}^2}{2m_y} $$
is the corresponding part of the electron energy. 

Equation (\ref{psitr1}) can be solved following the scheme developed for a constant magnetic filed \cite{BrizhikMF1}. In this way we can obtain for the unknown function the harmonic oscillator type equation 
 $\varphi (z, t)$ 
\begin{equation}
	\label{varphi}
	\biggl(
	-i\hbar \frac{\partial }{\partial t} + \frac{e^2 B_0^2 \cos^2 \omega t}{2 m_y c^2} (z - z_0)^2 
	-\frac{\hbar^2}{2m_z} \frac{\partial^2}{\partial z^2}  
	\biggr)
	\varphi (z, t) = E_{tr} \varphi (z, t),
\end{equation}
in which the notation is used
\begin{equation}
	\label{z0}
	{{z}_{0}}=-\frac{\hbar c{{k}_{y}}}{e{{B}_{0}}\cos \omega t}.
\end{equation}
In the quasi-stationary case $
\partial \varphi (z,t)/\partial t=0$ the solution of this equation is 
\begin{equation}
	\label{varphi1}
	\varphi ({z}')=exp\left( -\frac{1}{2}\frac{m_z{{\omega }_{0}}}{\hbar }{{{{z}'}}^{2}} \right){{H}_{n}}({z}'), \qquad {z}'=z-{{z}_{0}},
\end{equation}
where $H_{n}(u)$ is Hermite polynomial: 
\begin{equation}
	\label{Hn}
	H_{n}(u)=(-1)^{n}\exp (u^{2})\frac{d^n}{du^n}\exp (-u^{2}).
\end{equation}
Thus, the electron dynamics in a transverse plane is the dynamics of a quantum harmonic oscillator with the eigenfrequency ${{\omega }_{0}}$, which depends not only on the intensity of the magnetic field $B_0$, but also on its frequency $\omega $:
\begin{equation}
	\label{omega0}
	{{\omega }_{0}}=\frac{|{{B}_{0}}e\cos (\omega t)|}{m_{tr}c},
\end{equation}  
and the oscillator mass $m_{tr}$,
\begin{equation}
	\label{mtr}
m_{tr}=\sqrt{{{m}_{y}}{{m}_{z}}}.
\end{equation}

This defines the final expression for the transverse component of the electron wavefunction:  
\begin{equation}
	{{\psi }_{tr }}(y, z, t)
	=\frac{1}{\sqrt{l_y}}H_{n}(z-z_0)
e^{ \left[ ik_{y}y -\frac{m_z\omega _0}{2\hbar }(z-z_0)^{2}-\frac{i}{\hbar}E_{tr }t\right]},
\end{equation}
and, thus, electron dynamics in a parallel oscillating magnetic field is a composition of the "free" soliton coherent movement along  the molecular chain and electron cyclotron oscillations in the perpendicular plane with the cyclotron frequency (\ref{omega0}) and mass (\ref{mtr}).

\section{The case of the perpendicular magnetic field }
	
In this section we consider the case of the external harmonic magnetic filed, oriented perpendicular to the  molecular chain direction, for instance, when it is parallel to $y$-axis
\begin{equation}
	\label{B-tr}
	\vec{B}^{(p)}(t)=({0,{B}_{0}}\cos \omega t, 0), \qquad 	\vec{A}^{(p)}=(0, 0, -{{B}_{0}}x\cos \omega t, 0).
\end{equation}	 
where $\omega $ is the field oscillation frequency.	
The Hamiltonian of the system can be written now in the form $H^{(p)}= H_{s}^{(p)}+H_{tr }^{(p)}$, where 
\begin{equation}
	\label{4}
	H_{s}^{(p)}=-\frac{{{\hbar }^{2}}}{2{{m}_{x}}}\frac{{{\partial }^{2}}}{\partial {{x}^{2}}}+\frac{1}{2{{m}_{z}}}{{\left( \hbar {{k}_{z}}+\frac{e}{c}{{B}_{0}}x\cos \omega t \right)}^{2}}
	-\sigma \rho (x,t)
\end{equation}
\begin{equation}
	\label{Htrl}
	H_{tr }^{(p)}=\frac{{{\hbar }^{2}}}{2{{m}_{y}}}\frac{{{\partial }^{2}}}{\partial {{y}^{2}}}
	+\frac{{{\hbar }^{2}}}{2{{m}_{z}}}\frac{{{\partial }^{2}}}{\partial {{z}^{2}}}.
\end{equation}
Therefore, the electron wavefunction can be represented as the product of two components
\begin{equation}
	\label{psi-tr}
\psi ^{(p)} (\vec{r})=\psi^{(p)}_s(x,t) \psi ^{(p)}_{tr} (y,z,t),
\end{equation}
where the $x$-component of the wavefunction satisfies the nonlinear equation 
\begin{equation}
	\label{psi-perp}
	\left[ i\hbar \frac{\partial }{\partial t}+ \frac{\hbar^2}{2m_x}\frac{\partial ^2}{\partial x^2}+\sigma \rho(x,t)\right]\psi ^{(p)}_s (x,t)=\frac{1}{2m_x} \left[\hbar k_x + \frac{e}{c}B_0 x cos(\omega t)\right] \psi ^{(p)}_s (x,t),
\end{equation}
and $ \psi ^{(p)}_{tr}(y,z,t)$ is a normalized plane wave function in the $yz$-plane. 

It is convenient to introduce the dimensionless variables
\begin{equation}
	\label{adim-par}
	\tilde{x}=\frac{\sqrt{g}}{a}x, \quad
	{{\tilde{x}}_{0}}=\frac{\sqrt{g}\hbar c{{k}_{z}}}{ae{{B}_{0}}}, \quad 
	\tau =2\frac{Jg}{\hbar }t,\quad \Omega =\frac{\hbar \omega }{2Jg},
\end{equation}
and to re-write Eq.(\ref{psi-perp}) in the form of the  soliton equation modified with the term in the right hand side:
\begin{equation}
	\label{6}
	\left[ i\frac{\partial }{\partial \tau }+\frac{1}{2}\frac{{{\partial }^{2}}}{\partial {{{\tilde{x}}}^{2}}}+\frac{\sigma \rho }{2Jg} \right]{\psi ^{(p)}_s }(\tilde{x},\tau )
	=\varepsilon {{\left( \tilde{x}\cos \Omega \tau +{{{\tilde{x}}}_{0}} \right)}^{2}}{{\psi ^{(p)}_s }}(\tilde{x},\tau ).
\end{equation}
Here the coefficient $\varepsilon$ is introduced,
\begin{equation*}
	\varepsilon =\frac{{{e}^{2}}{{B}_{0}}^{2}{{a}^{2}}}{4{{m}_{z}}J{{c}^{2}}{{g}^{2}}}
\end{equation*}
which is small even for very strong magnetic fields, $\varepsilon \ll 1$.

In view of this, the chain deformation $\rho (x, t)$ can be represented as the sum of two terms	
\begin{equation}
	\label{7}
	\rho (x,t)={{\rho }_{0}}(u)+\varepsilon {{\rho }_{1}}(u),
\end{equation}
where the running wave variable is introduced
\begin{equation*}
	u=x-\xi (t).
\end{equation*}
Substituting expression (\ref{7}) into Eq. (\ref{eqrho1}), we get 
\begin{equation}
	\label{8}
	{\rho }_{0}(u)=\frac{\sigma }{w  (1-s^{2}) }{{\left| \psi^{(p)} _s(u) \right|}^{2}}.
\end{equation}
Here $s^2=V^2/V_0^2 $ as in the absence of the magnetic field, but the soliton velocity $V$ is defined now in the general form
\begin{equation*}
	\label{v}
	V=\frac{d\xi }{dt}.
\end{equation*}
The second component of the deformation, ${{\rho }_{1}}(u)$, in Eq. (\ref{7}) satisfies the equation
\begin{equation}
	\label{9}
	\frac{d{{\rho }_{1}}}{du}=-\frac{V'}{\varepsilon {{V}_{0}}^{2}\left( 1-s^{2} \right)}{{\rho }_{0}},
\end{equation}
where $V'=dV/dt $ is the acceleration of the soliton. 

Substituting $\rho (x, t)$ in Eq. (\ref{6}), we get the NLS with the extra term in the right hand side
\begin{equation}
	\label{10}
	\left[ i\frac{\partial }{\partial \tau }+\frac{1}{2}\frac{{{\partial }^{2}}}{\partial {{{\tilde{x}}}^{2}}}+\frac{1}{2}|{{\psi }_{s}^{(p)}}{{|}^{2}} \right]{{\psi }_{s}^{(p)}}
	=i\varepsilon R[\psi_{s}^{(p)} ],
	\end{equation}
where ${\psi }_{s}^{(p)}= {{\psi }_{s}^{(p)}}(\tilde{x},\tau )$ and 
\begin{equation}
	\label{12}
	R[\psi _{s}^{(p)}]=-i\left[ {{\left( \tilde{x}\cos \Omega \tau +{{{\tilde{x}}}_{0}} \right)}^{2}}-\frac{\sigma }{2Jg}{{\rho }_{1}}(\tilde{x},\tau ) \right]\psi _{s}^{(p)}.
\end{equation}

Due to the presence of the term $ R[\psi _{s}^{(p)}]$ in Eq. (\ref{10}), soliton parameters are functions of time. Since the coefficient $\varepsilon$ is small,  this dependence is weak and equations which determine this dependence, can be obtained using the perturbation method, based on the  inverse scattering technique for the NLS \cite{Karpman}. Thus, we search the solution of Eq. (\ref{10}) in the form
\begin{equation}
	\label{13}
	{{\psi }_{s}^{(p)}}(\tilde{x},\tau )=2\nu \operatorname{sech}\zeta \exp \left[ i\varphi \right]
\end{equation}
where the coordinate and phase are introduced:
\begin{equation*}
	\zeta =2\nu \left( \tilde{x}-\tilde{\xi } \right), \qquad
	\varphi =\frac{\mu \zeta }{\nu }+\eta .
\end{equation*} 

Substituting now expression (\ref{13}) into Eq. (\ref{8}) and solving Eq. (\ref{9}), we get the solution for the term of the chain deformation ${{\rho }_{1}}$, arising due to the perturbation by the magnetic field :
\begin{equation}
	\label{rho1}
	{{\rho }_{1}}=-\frac{4\nu \sigma aV'}{\varepsilon \sqrt{g}{{V}_{0}}^{2}\kappa ( 1-s^{2})}th\left[ 2\nu \left( \tilde{x}-\tilde{\xi } \right) \right].
\end{equation}

Thus, Eq. (\ref{12}) takes the form
\begin{equation}
	\label{R}
	R[\psi _s^{(p)}]=-2i\nu sech \zeta \biggl[ {{\left( \tilde{x}\cos \Omega \tau +{{{\tilde{x}}}_{0}} \right)}^{2}}+ \frac{2\nu {{\sigma }^{2}}aV'th\zeta }{\varepsilon Jg^{3/2}V_{0}^{2}w ( 1-s^{2})}
	\biggr]\exp \{i\varphi \}.
	\end{equation}

According to the perturbation method \cite{Karpman}, soliton parameters $\nu$, $\mu$, $\tilde{\xi }$, $\eta$ are determined by the following equations
\begin{equation*}
	\frac{d\nu }{d\tau }=\frac{\varepsilon }{2}\operatorname{Re}\int\limits_{-\infty }^{\infty }{\operatorname{sech}\zeta R[\psi _s^{(p)}]\exp (-i\varphi )d\zeta },
\end{equation*}
\begin{equation*}
	\frac{d\mu }{d\tau }=\frac{\varepsilon }{2}\operatorname{Im}\int\limits_{-\infty }^{\infty }{\frac{\operatorname{sh}\zeta }{c{{h}^{2}}\zeta }R[\psi ]\exp (-i\varphi )d\zeta },
\end{equation*}
\begin{equation*}
	\frac{d\tilde{\xi }}{d\tau }=-\frac{1}{2\nu }Im\left( -2{{\left( \mu +i\nu  \right)}^{2}} \right)+
	+\frac{\varepsilon }{4{{\nu }^{2}}}\operatorname{Re}\int\limits_{-\infty }^{\infty }{\frac{\zeta }{ch\zeta }R[\psi ]\exp (-i\varphi )d\zeta },
\end{equation*}
\begin{equation*}
	\frac{d\tilde{\xi }}{d\tau }=2\mu \frac{d\tilde{\xi }}{d\tau }+\operatorname{Re}\left( -2{{\left( \mu +i\nu  \right)}^{2}} \right)
	+\frac{\varepsilon }{2\nu }\operatorname{Im}\int\limits_{-\infty }^{\infty }{\frac{1-\zeta th\zeta }{ch\zeta }R[\psi ]\exp (-i\varphi )d\zeta }.
\end{equation*}
Solving these equations, we get the time dependence of the soliton parameters:
\begin{equation*}
	\nu ={{C}_{1}}, 
\end{equation*}	
\begin{equation*}
	\tilde{\xi }=2\mu \tau +{{C}_{2}}, 
\end{equation*}	
\begin{equation*}
	\mu =-\varepsilon \biggl( \frac{{\tilde{\xi }}}{2\Omega }\left( \Omega \tau +\sin \Omega \tau \cos \Omega \tau  \right)-
	-\frac{{{{\tilde{x}}}_{0}}}{\Omega }\sin \Omega \tau +\frac{2}{3}\nu \tau \alpha  \biggr)+{{C}_{3}}, 
\end{equation*}	
\begin{equation*}
	\eta =2\tau \left( {{\nu }^{2}}+{{\mu }^{2}} \right)+\varepsilon \left( \frac{{{\pi }^{2}}}{48\Omega {{\nu }^{2}}}-\frac{{{{\tilde{\xi }}}^{2}}}{\Omega } \right)\biggl( \frac{\Omega \tau }{2}
	+\frac{1}{2}\sin \Omega \tau \cos \Omega \tau  \biggr)+\frac{2\varepsilon \tilde{\xi }{{{\tilde{x}}}_{0}}}{\Omega }\sin \Omega \tau -\varepsilon \tau {{\tilde{x}}_{0}}^{2}+{{C}_{4}} ,
\end{equation*}
where ${{C}_{1}}$, ${{C}_{2}}$, ${{C}_{3}}$, ${{C}_{4}}$ are constants of integration, and parameter $\alpha $ is introduced:
\begin{equation*}
\alpha =\frac{2\nu {{\sigma }^{2}}aV'}{\varepsilon w J{{g}^{3/2}}({{V}_{0}}^{2}-{{V}^{2}})}
\end{equation*}

Finally, we obtain the soliton wave function  
\begin{equation}\label{psisl}
{{\psi }_{s}^{(p)}}(\tilde{x},\tau )=2\nu \operatorname{sech} {\left( 2\nu \left( \tilde{x}-\tilde{\xi } \right)\right) } \exp \left[ {2i\mu \left( \tilde{x}-\tilde{\xi } \right) }+i\eta \right]
\end{equation}
in which the explicit expressions for depending on time soliton  parameters, taking into account the normalization condition, are given below
\begin{equation}
\nu ={{C}_{1}}, 
\end{equation}
\begin{equation}
	\label{ksi}
\tilde{\xi }=\frac{2\tau \varepsilon \left( \frac{{{{\tilde{x}}}_{0}}}{\Omega }\sin \Omega \tau -\frac{2}{3}\tau \alpha {{}_{1}} \right)+2\tau {{C}_{3}}+{{C}_{2}}}{1+\frac{\varepsilon \tau }{\Omega }\left( \Omega \tau +\sin \Omega \tau \cos \Omega \tau  \right)}, 
\end{equation}
\begin{multline}
	\label{mu}
\mu =-\varepsilon \biggl[ \frac{\frac{\tau \varepsilon }{\Omega }\left( \frac{{{{\tilde{x}}}_{0}}}{\Omega }\sin \Omega \tau -\frac{2}{3}\tau \alpha {{}_{1}} \right)+\tau {{C}_{3}}+\frac{{{C}_{2}}}{2}}{1+\frac{\varepsilon \tau }{\Omega }\left( \Omega \tau +\sin \Omega \tau \cos \Omega \tau  \right)}\biggl( \Omega \tau +
+\sin \Omega \tau \cos \Omega \tau  \biggr)-\\
\frac{{{{\tilde{x}}}_{0}}}{\Omega }\sin \Omega \tau +\frac{2}{3}{{C}_{1}}\tau \alpha  \biggr]+{{C}_{3}}, 
\end{multline}
\begin{multline}
	\label{eta}
\eta =2\tau \left( {{\nu }^{2}}+{{\mu }^{2}} \right)+\varepsilon \left( \frac{{{\pi }^{2}}}{48\Omega {{\nu }^{2}}}-\frac{{{{\tilde{\xi }}}^{2}}}{\Omega } \right)\biggl( \frac{\Omega \tau }{2}+\\
+\frac{1}{2}\sin \Omega \tau \cos \Omega \tau  \biggr)+\frac{2\varepsilon \tilde{\xi }{{{\tilde{x}}}_{0}}}{\Omega }\sin \Omega \tau -\varepsilon \tau {{\tilde{x}}_{0}}^{2}+{{C}_{4}}
\end{multline}
and remind that here the dimensionless parameters and frequency $\Omega $ are determined in Eq. (\ref{adim-par}).

%\section{Role of energy dissipation}
	
%In real systems there is always present dissipation of energy. Very often energy dissipation plays negative role. But the situation is very different when we consider nonlinear systems, such as our case. 

	\section{Conclusions}

The obtained results show that the impact of the external oscillating in time magnetic field on soliton dynamics has a complex nature and  essentially depends on the orientation of the field with respect to the orientation of a molecular chain. In the presence of a longitudinal magnetic field, the soliton propagates along the molecular chain as a free soliton, and its dynamics in the perpendicular plane is the oscillatory-type movement, described by functions of the harmonic oscillator with the cyclotron frequency  
 depending both on the intensity and frequency of the magnetic field according to Eq. (\ref{omega0}). In a perpendicular to chain field, soliton propagates along the chain with the velocity and phase, depending on time, while the perpendicular to chain orientation component of the soliton wavefunction is represented by the normalized plane wave. According to solutions (\ref{ksi})-(\ref{eta}), soliton parameters essentially depend on time and their dependence is given by complex functions which can be expanded in series that includes harmonics with the main frequency of the external field and its higher harmonics. It is worth to mention that in the present study we have used adiabatic approximation and represented the chain deformation in the form (\ref{8}), proportional to the electron wavefunction in the zero-order approximation and neglected the impact of the higher order corrections of the wavefunction. We expect, that the proper account of these higher order corrections will modify the deformation, and, as a result, will lead to 
 quasi-resonant dynamic response of solitons to the magnetic field due to the 
 radiation of the linear waves (cf. with soliton radiation in the electromagnetic field \cite{BC-HE}). This work is in progress.

Thus, we conclude that such complex impact of the magnetic field on soliton dynamics essentially affects the conductivity of low-dimensional systems with intermediate values of the electron-lattice interaction, whose ground electron state corresponds to soliton. Hence, we can predict that magnetic fields can significantly affect charge transport processes in micro- and nanoelectronic devices, and the devices of biomimetic technologies that are based on polymers and other quasi-one-dimensional materials (see, e.g., \cite{Sirri,Yao,Shelte,Ahmad} and references therein), as well as therapeutic and diagnostic devices based on the so called "next generation" low-dimensional materials (see \cite{Minamiki,Chen2,Zhou}).

Another important conclusion which follows from the present study, concerns biological systems: we have shown that external alternating magnetic fields, changing the dynamics of solitons, affect charge and energy transport in the redox processes in biological systems, and, thus, we can expect that the obtained results can explain the physical mechanism of the therapeutic effects of low intensity oscillating magnetic fields \cite{LiboffSmith,BrizhikFermi,Buchachenko,Emre,Funk}.

	\vskip5mm 
	{\bf Acknowledgement.} 
	\textit{The author expresses thanks to K. Temchenko for fruitful discussions during her work on the master degree thesis. This work was supported by the fundamental scientific program 0122U000887 of the Department of Physics and Astronomy of the National Academy of Sciences of Ukraine. The author acknowledges also the Simons Foundation.}


\begin{thebibliography}{99} 
	
\bibitem{DavydovKislukha1}
A.S. Davydov, N.I. Kislukha, Solitary excitons in one-dimensional molecular chains. Phys. Status Solidi B \textbf{59}, 465 (1973).  https://doi.org/10.1002/pssb.2220590212
\bibitem{Davydov}
A.S.Davydov, \textit{Solitons in Molecular Systems}, Dordrecht, Reidel, 1991.
\bibitem{Scott}
A. C. Scott, Davydov's soliton. Phys. Rep. \textbf{217}, 1 (1992).
\bibitem{Brizhik-dyn}
L.S. Brizhik. Dynamical properties of Davydov solitons. Ukr. J. Phys.,   \textbf{48},  611-622 (2003).
\bibitem{Brizhik-DAA}
L.S. Brizhik, J. Luo, B.M.A.G. Piette, W.J. Zakrzewski. Long-range electron transport mediated by alpha-helices, Phys. Rev. E, \textbf{100}  062205  (2019).  DOI: 10.1103/PhysRevE.100.062205.
\bibitem{BrizhikMF1}
 L.S. Brizhik. Soliton dynamics in constant magnetic field, 
 %Theor. Mat. Fiz., 1990, 83, No.3, 342-347; 
 Theor. Math. Phys \textbf{83} (1990): 578-582.
\bibitem{BrizhikMF2}
L.S. Brizhik. Bisoliton in constant magnetic field, Phys. Stat. Sol.(b), 1990, \textbf{157}, No.2, 649-655.
\bibitem{BrizhikArXiv}
L. Brizhik, Biological effects of pulsating magnetic fields: role of solitons. http://arxiv.org/abs/1411.6576
\bibitem{BrizhikFermiZavan}
L. Brizhik, E. Fermi, B. Zavan. Working principle of magnetic resonance therapy. J. Adv. Phys. 2015; http://arxiv.org/abs/1509.04475
\bibitem{BrizhikFermi}
L. Brizhik, L. Ferroni, C. Gardin, E. Fermi. On the Mechanisms of Wound Healing by Magnetic Therapy: the Working Principle of Therapeutic Magnetic Resonance. Int. J. Biophys, 2016, \textbf{6}, pp. 27-43; DOI: 10.5923/j.biophysics.20160603.01.
\bibitem{McLeod} 
B.R. McLeod, A.R. Liboff. Cyclotron resonance in cell membrane: the theory of the mechanisms. In: Mechanistic Approaches to Interactions of Electric and Electromagnetic Fields with Living Systems. Blank M., Findl E. (eds). New York and London: Plenum Press, (1987) pp. 97–108.
\bibitem{TemchBrizhik} 
K.V. Temchenko, L.S. Brizhik. On the dynamics of Davydov's soliton in scillating magnetic field oriented along the molecular chain. Report at the XIV scientific conference of students, PhD students and young scientists of the Physics-Technical Institute of the National Technical University of Ukraine “Igor Sikorsky Kyiv Polytechnic Institute”, 2016.
%Темченко К. В., Брижик Л. С. Динамiка давидовського солiтона в осцилюючому магнiтному полi, спрямованому вздовж молекулярного ланцюжка — Труди XIV Наукової конференції студентів, аспірантів та молодих вчених на базі Фізико-технічного інституту НТУУ «КПІ», 2016.
%\bibitem{BrizhikTemch2}  
%Tемченко К. В., Брижик Л. С. Вплив осцилюючого магнiтного поля, перпендикулярного молекулярним ланцюжкам, на динамiку молекулярних солiтонiв — Труди XIV Наукової конференції студентів, аспірантів та молодих вчених на базі Фізико-технічного інституту НТУУ «КПІ», 2016.
\bibitem{Karpman}
Karpman V. I., Maslov E. M. Perturbation theory for solitons. JETP  \textbf{73} (1977) 537-559.
\bibitem{BC-HE}
L.Brizhik, L.Cruzeiro-Hansson, A.Eremko. Influence of electromagnetic radiation on  molecular solitons, J. Biol. Physics 1998, v. 24, No. 1, 19-39.
\bibitem{Sirri}
H. Sirringhaus. 25th Anniversary article: Organic field-effect transistors: the path beyond amorphous silicon. Adv. Mater. 26 (2014) 1319.
\bibitem{Yao}
Z.F. Yao, J.Y. Wanga and J. Pei. High-performance polymer field-effect transistors: from the perspective of multi-level microstructures. Chem. Sci., \textbf{12} (2021) 1193. doi.org/10.1039/D0SC06497A
\bibitem{Shelte}
A.R. Shelte, S. Pratihar. Chapter 14. Next-generation nanomaterials for
environmental industries: Prospects and challenges.  
Green Functionalized Nanomaterials for Environmental Applications. Micro and Nano Technlogies 2022, pp. 399-415. https://doi.org/10.1016/B978-0-12-823137-1.00015-4
\bibitem{Ahmad}
Z. Ahmad, M.K. Abdullah, M.Z. Ali, M.A. Md Zawawi. Polymers in Electronics. Elsevier, 2023. ISBN 9780323983822, https://doi.org/10.1016/B978-0-323-98382-2.01001-2.
\bibitem{Minamiki}
T. Minamiki, Y. Sasaki, S. Su and T. Minami. Development of polymer field-effect transistor-based immunoassays. Polymer J. \textbf{51} (2019) 1-9. https://doi.org/10.1038/s41428-018-0112-0
\bibitem{Chen2} 
Z. Chen, Z. Wang, and Z. Gu. Bioinspired and Biomimetic Nanomedicines. Acc. Chem. Res. \textbf{52} (2019) 1255. https://doi.org/10.1021/acs.accounts.9b00079
\bibitem{Zhou}
J. Zhou,  L. Zhang. Biomimetic nanotechnology toward personalized vaccines
Adv. Mater., \textbf{32} (2020), 1901255. %https://doi.org/10.1038/s41428-018-0112-0
[immune] Zhang, H.; Ye, H.; Zhang, Q.; Zeng, F.; Huang, X.; Zhang, Q.; Li, Z.; Du, B. Experimental studies on extremely low frequency pulsed magnetic field inhibiting sarcoma and enhancing cellular immune functions. Sci China C Life Sci 1997, 40, 392-397.
\bibitem{LiboffSmith}
A.R. Liboff, S.D. Smith, B.R. McLeod. Experimental evidence for ion cyclotron resonance mediation of membrane transport. In: Mechanistic Approaches to Interactions of Electric and Electromagnetic Fields with Living Systems. Blank M., Findl E. (eds). New York and London: Plenum Press, (1987) pp. 109–132.
\bibitem{Buchachenko}
 Buchachenko, A.; Kouznetsov, D. Magnetic field affects enzymatic ATP synthesis. J Am Chem Soc 2008, 130, 12868-12869.
\bibitem{Emre}
%[oxidative-stress] 
Emre, M.; Cetiner, S.; Zencir, S.; Unlukurt, I.; Kahraman, I.; Topcu, Z. Oxidative Stress and Apoptosis in Relation to Exposure to Magnetic Field. Cell Biochem Biophys 2011, 59, 71-77.
\bibitem{Funk}
R.H. Funk. Coupling of pulsed electromagnetic fields (PEMF) therapy to molecular grounds of the cell. Am. J. Transl. Res. 10(5) (2018) 1260-1272. eCollection 2018. Review.


\end{thebibliography}
\end{document}